\documentclass[twocolumn,showpacs,preprintnumbers,amsmath,amssymb,superscriptaddress,aps,pra,10pt]{revtex4-1}

\usepackage{color}
\usepackage{graphicx}
\usepackage{citesort}
\usepackage{color}
\usepackage{subfigure}
\usepackage{amsmath,amssymb}
\usepackage{upgreek}

\newcommand{\comment}[1]{}

\begin{document}

\title{Coherently refreshed acoustic phonons for extended light storage
 }

\author{Birgit Stiller$^{1,2,3,\ast,\dagger}$, Moritz Merklein$^{1,2,\ast}$, Christian Wolff$^{3}$, Khu Vu$^{4}$,  Pan Ma$^{4}$, Stephen J. Madden$^{4}$ and Benjamin J. Eggleton$^{1,2}$\\
\small{ \textcolor{white}{blanc\\}
$^{1}$Institute of Photonics and Optical Science (IPOS), School of Physics, University of Sydney, Sydney, NSW, 2006 Australia.\\
$^{2}$The University of Sydney Nano Institute (Sydney Nano), University of Sydney, Sydney, NSW, 2006 Australia.\\
$^{3}$Max-Planck-Institute for the Science of Light, Staudtstr. 2, 91058 Erlangen, Germany.\\
$^{4}$Center for Nano Optics, University of Southern Denmark, Campusvej 55, DK-5230 Odense M, Denmark.\\
$^{5}$Laser Physics Centre, RSPE, Australian National University, Canberra, ACT 0200, Australia.\\
$^{\ast}$These authors contributed equally to this work.\\
$^{\dagger}$birgit.stiller@mpl.mpg.de}}

\begin{abstract}

Acoustic waves can serve as memory for optical information, however, acoustic phonons in the GHz regime decay on the nanosecond timescale. Usually this is dominated by intrinsic acoustic loss due to inelastic scattering of the acoustic waves and thermal phonons. Here we show a way to counteract the intrinsic acoustic decay of the phonons in a waveguide by resonantly reinforcing the acoustic wave via synchronized optical pulses. This scheme overcomes the previous constraints of phonon-based optical signal processing for light storage and memory. We experimentally demonstrate on-chip storage up to 40\,ns, four times the intrinsic acoustic lifetime in the waveguide. We confirm the coherence of the scheme by detecting the phase of the delayed optical signal after 40\,ns using homodyne detection. Through theoretical considerations we anticipate that this concept allows for storage times up to microseconds within realistic experimental limitations while maintaining a GHz bandwidth of the optical signal. The refreshed phonon-based light storage removes the usual bandwidth-delay product limitations of e.g. slow-light schemes.

\end{abstract}

\maketitle


Coupling optical and mechanical waves in cavities, waveguides and nanostructures offers great potential for optical signal processing \cite{Safavi-Naeini2011b,Pant2011,Rakich2013,Eggleton2013,Beugnot2014,VanLaer2015,Balram2015,Li2015a}, in particular for delay lines and storage schemes \cite{Zhu2007,Chang2010a,Safavi-Naeini2011,Fiore2011,Jamshidi2012,Fiore2013,Galland2014,Dong2015,Merklein2017,Merklein2018,Stiller2019}. One particular optomechanic interaction with GHz acoustic phonons is Brillouin scattering, which describes the interaction between optical and traveling acoustic waves \cite{Brillouin1922,Bloembergen1965}. While spontaneous scattering is initiated by noise \cite{Boyd1990,Gaeta1991}, and hence is not coherent, stimulated Brillouin scattering (SBS) involves a coherent excitation of acoustic phonons with GHz frequency solely by optically forces. It was shown recently that one can use these acoustic phonons to store and delay optical signals \cite{Zhu2007,Merklein2017,Stiller2019}. The optical information is resonantly transferred to a coherent acoustic phonon and is then transferred back to the optical domain by a delayed optical retrieval pulse completely preserving the phase and amplitude \cite{Merklein2017} and the wavelength of the signal \cite{Stiller2019}. However, these high-frequency acoustic phonons decay exponentially with a lifetime of a few nanoseconds determined by the material properties at room temperature. This inherent decay is due to the damping of the acoustic waves while propagating through the material. Therefore, the optical information stored in the acoustic waves is lost and a way of preserving the coherent acoustic vibration is needed.

\begin{figure*}[t]
  \centering
  \includegraphics[width=0.94\textwidth]{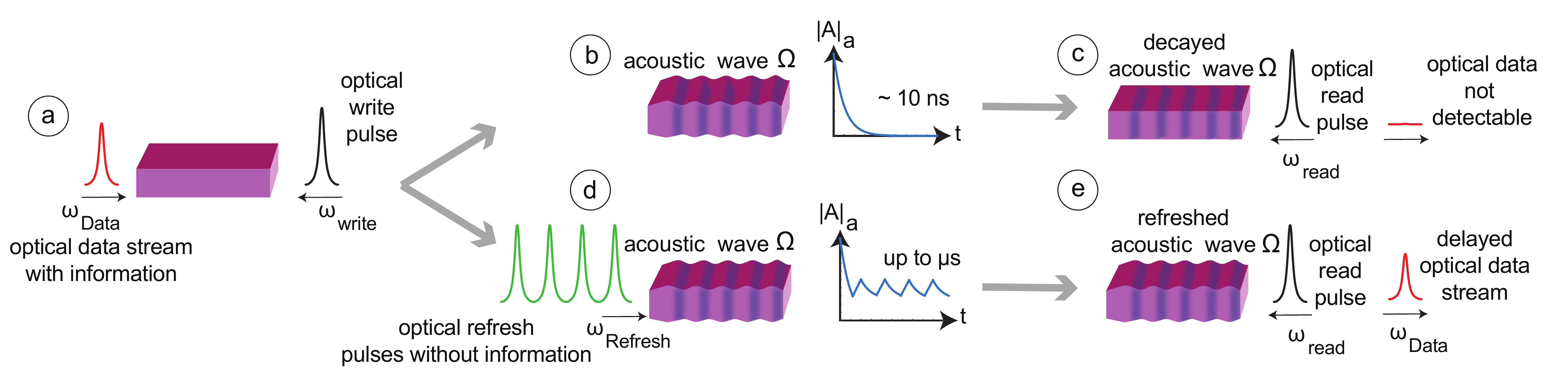}
\caption{Concept of the refreshed acoustic memory. a) An optical write pulse converts the information of an optical data stream to an acoustic wave; b) The acoustic wave propagates at a 5 orders of magnitude lower speed in the waveguide and decays with the acoustic lifetime; c) in normal operation the acoustic wave dissipates and the read pulse cannot sufficiently interact with the acoustic wave; therefore the information of the optical data is lost; d) to counteract the acoustic decay, optical refresh pulses at $\omega_{\mathrm{refresh}}$ = $\omega_{\mathrm{data}}$ transfer energy to the acoustic phonons; e) an optical read pulse converts the information back to the optical domain and the delayed optical information exits the waveguide.}
\label{fig1}
\end{figure*}

Here, we introduce and demonstrate a concept to counteract the intrinsic acoustic decay of the phonon in a waveguide by resonantly reinforcing the coherent acoustic phonons via synchronized optical pulses. Instead of converting the acoustic waves back to the optical domain, refresh pulses at the wavelength of the original data pulses transfer energy to the acoustic wave and counteract the decay. We experimentally demonstrate that information can be stored and retrieved for 40\,ns - a time much longer than the intrinsic acoustic lifetime of 10\,ns - and we confirm that the coherence is preserved in this process by measuring the optical phase after 40\,ns via homodyne detection. We also experimentally demonstrate an increase in readout efficiency for storage times shorter than the acoustic lifetime. We theoretically explore the limits of the scheme and demonstrate that within practical limits even storage times up to micro seconds are within reach while maintaining a broad GHz bandwidth of the stored optical pulses. This scheme allows the extension of the lifetime of the coherent acoustic phonons and the efficiency of optoacoustic memory, removing previous constraints of phonon-based optical signal processing schemes. Most importantly, it decouples the possible delay time from the bandwidth of the stored pulses, which typically limits slow light schemes based on nonlinear effects, atomic vapours and cold atoms.


The storage concept is based on the coupling of two optical waves with a traveling acoustic wave via the effect of stimulated Brillouin scattering. The acousto-optic coupling depends on the overlap of the optical and acoustic waves and the intrinsic photo-elastic material response. The addressed acoustic wave is at $\Omega=$7.8\,GHz and the nonlinear gain is in the order of G$_b=$500\,W$^{-1}$m$^{-1}$ for the used photonic waveguide made out of As$_2$S$_3$ with a cross-section of 2.2\,$\upmu$ by 800\,nm.
Figure 1 shows schematically the principle of storing, coherently refreshing the acoustic phonons and retrieving the delayed optical information. The information of the optical data pulse is initially transferred to the acoustic phonons by a counter-propagating optical write pulse, offset in frequency by the acoustic resonance frequency of the waveguide $\omega_{\mathrm{acoustic}}$ = $\omega_{\mathrm{write}}$ - $\omega_{\mathrm{data}} = $7.8\,GHz (Fig. 1a and b). The acoustic resonance frequency relates to the acoustic velocity $V_{A}$, the effective refractive index $n_{\mathrm{eff}}$ and the pump wavelength $\lambda_{\mathrm{Pump}}$ as $\Omega=\frac{2 n_{\mathrm{eff}}V_{A}}{\lambda_{\mathrm{Pump}}}$. The acousto-optic coupling not only requires energy conservation but also phase-matching as $k_{\mathrm{acoustic}}$ = $k_{\mathrm{write}}$ - $k_{\mathrm{data}}$. An efficiency of up to 30\,$\%$ can be reached depending on the bandwidth of the optical pulse \cite{Merklein2017}. This initial storage process is a Anti-Stokes process because the data wave looses energy to the acoustic wave. Without further action, the acoustic phonons decay after several nanoseconds (Fig. 1b) due to the intrinsic dissipation of the material. A read pulse at $\omega_{\mathrm{read}}$ = $\omega_{\mathrm{write}}$ cannot efficiently couple to the acoustic grating and the information is lost (Fig. 1c). To reinforce the acoustic vibration, we use optical refresh pulses (Fig. 1d) with the same frequency and propagation direction as the data pulse, $\omega_{\mathrm{refresh}}$ = $\omega_{\mathrm{data}}$. Herewith, the refresh pulses are scattered by the existing acoustic phonons - with two consequences: a portion of energy is transferred to the acoustic phonons, which refreshes the memory, and pulses with less energy at frequency $\omega_{\mathrm{write}}$ = $\omega_{\mathrm{refresh}}$ - $\omega_{\mathrm{acoustic}}$ are backscattered. This can be related to a Stokes process originating from the refresh pulses and the coherent acoustic wave. In order to retrieve the original data, a counter-propagating optical read pulse at $\omega_{\mathrm{read}}$ = $\omega_{\mathrm{write}}$ finally converts the information stored in the acoustic phonons back to the optical domain (Fig. 1e).

\begin{figure}[b]
  \centering
  \includegraphics[width=0.45\textwidth]{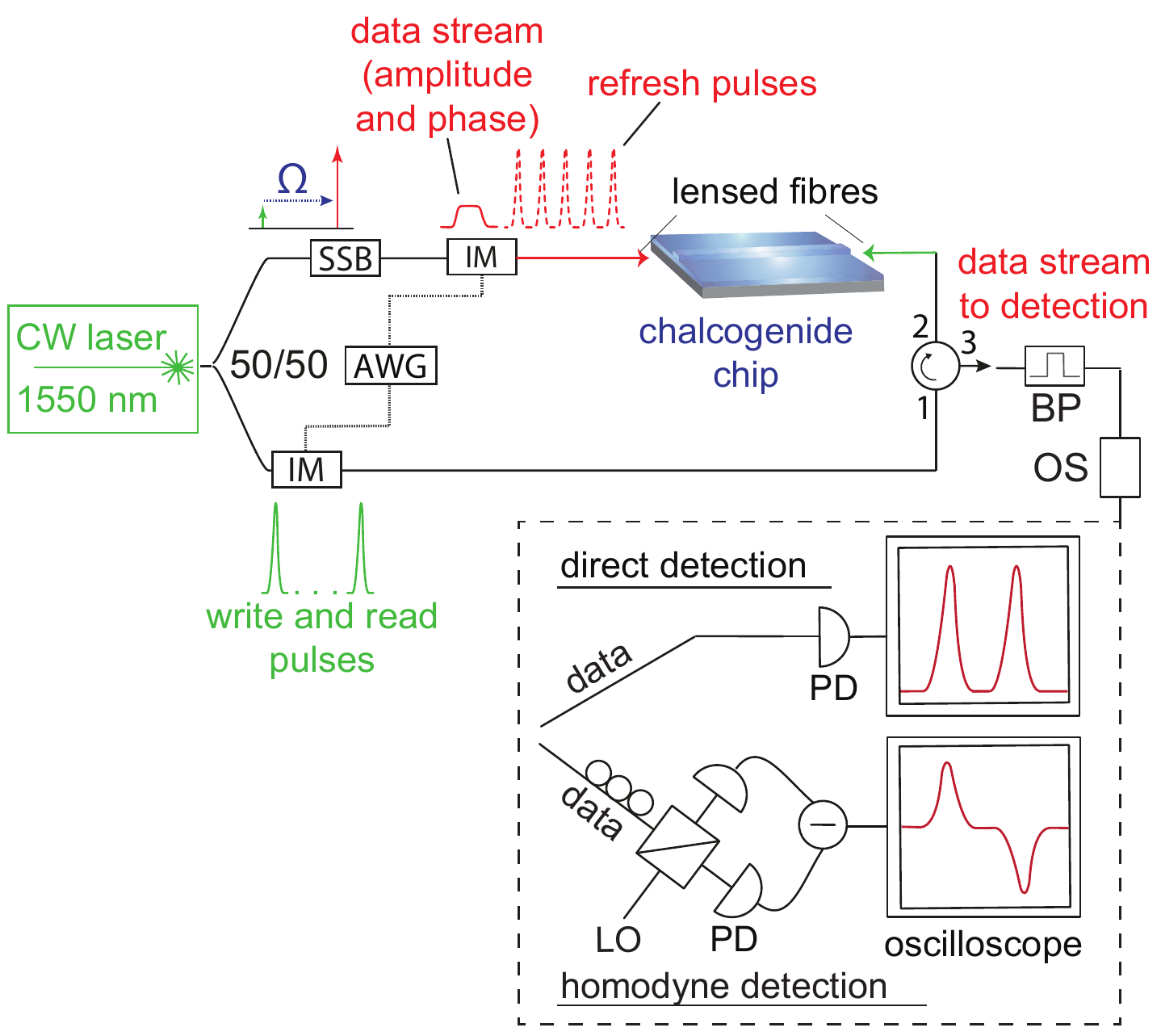}
\caption{Experimental setup for the refreshed optoacoustic memory. SSB single-sideband modulator, IM intensity modulator, AWG arbitrary waveform generator, BP bandpass filter, OS optical switch, LO local oscillator, PD photo diode.}
\label{figsetup}
\end{figure}

The refresh process can also be understood in the context of coherent heating \cite{Townes1963} as the existing acoustic phonons are coherently amplified by the refresh pulses which satisfy the energy and momentum requirements for a Stokes process. It can also be seen as a classical SBS backscattering process, however not initiated by thermal phonons but initiated by a deterministic localized seed created through the previous storage process. The refresh pulses do not contain any information but are coherent and synchronized with the data pulse. The number of refresh pulses depends on how long the storage is needed and in principle can extend the memory by several orders of magnitudes, fully countervailing the intrinsic exponential decay of the acoustic wave. However, in practice the signal-to-noise ratio (SNR) of the optical pulses, the dissipation of the material at room temperature and the broadening of the acoustic dynamic grating due to the convolution with finite control pulses limit the time after which the delayed optical signal can still be detected.

\begin{figure}[t]
   \centering
   \includegraphics[scale=0.55]{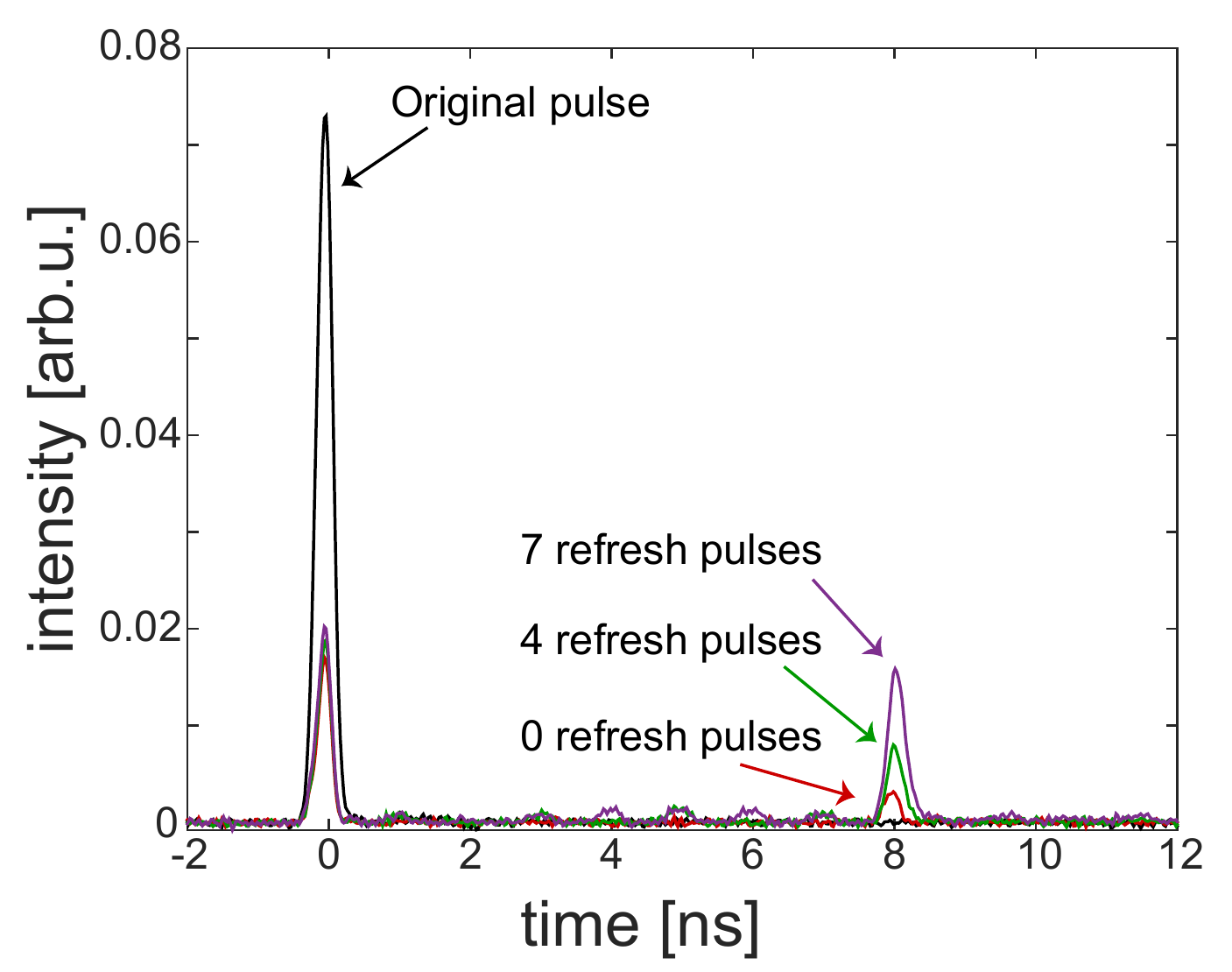}
 \caption{Experimental results for the refreshed Brillouin-based memory with a read-out after 8\,ns (within the acoustic lifetime): comparison of the efficiency while using 0, 4 and 7 refresh pulses showing a three and five times enhancement, respectively.}
 \label{fig2}
\end{figure}

The refreshed optoacoustic memory is implemented in a highly nonlinear As$_2$S$_3$ chip with 2.2$\upmu$m-large rib waveguides. A simplified experimental setup is shown in Fig. 2. A continuous wave diode laser at 1550\,nm is separated into two branches: one for the write and read pulses and the other one for the data stream and refresh pulses. The write and read pulses with 500\,ps duration are carved in by an electro-optic modulator driven by an arbitrary waveform generator. The time distance between them can be adjusted arbitrarily and defines the storage time in the memory. They are amplified to about 20\,W peak power and coupled into the chip from one side by lensed fibers. The other optical branch is firstly upshifted in frequency by the corresponding Brillouin frequency shift, here 7.8\,GHz. The frequency shift is implemented by a single-sideband modulator. Then a data stream in amplitude and phase is encoded by a second electro-optic modulator and a second channel of the arbitrary waveform generator. The data pulses are 500\,ps long and are amplified to a peak power of about 100\,mW and inserted from the opposite site as the write and read pulses into the photonic chip. In order to reinforce the acoustic wave, coherent refresh pulses are sent into the photonic waveguide, following the data stream. Here, we experimentally implement the refresh pulses by the same modulator as the data stream. The pulses are 300\,ps long and the peak power is varied to match the appropriate pulse area \cite{Zhu2007,Winful2015}, here about 200\,mW. The length, peak power and number of the refresh pulses have to be adjusted carefully in order to minimize distortion due to spontaneous Brillouin scattering and interaction with the optical background of the write and read pulses. After passing through the photonic chip, the data stream is filtered by a 3\,GHz broad filter to prevent from detecting backreflections of the write and read pulses at another wavelength. Then an optical switch filters out residual refresh pulses. This optical switch can be made superfluous when using the opposite polarization for refreshing the memory. The detection is done either directly by a photodiode (amplitude) or using homodyne detection (phase). In the latter case, the data is interfered with a local oscillator at the same wavelength and the different phases are seen as positive or negative signal on the oscilloscope.


As a first experimental proof, we show that the efficiency of the Brillouin-based storage increases when the acoustic wave is refreshed (Fig. 2). Therefore, we compare the amplitude of the retrieved data at a given storage time of 8\,ns. Without refresh pulses, the retrieval efficiency is about 4\,$\%$. With 4 or 7 refresh pulses, the efficiency can be increased to 10\,$\%$ and 20\,$\%$, respectively. The 7 refresh pulses were sent in with a time delay of 1\,ns after the data pulse with 1\,ns time spacing, the 4 refresh pulses at times 1, 3, 5 and 7\,ns after the data pulse. We use an optical switch to remove residual refresh pulses. However, as an alternative, the refresh pulses could also be inserted at orthogonal polarization, such that an optical switch is not necessary.

\begin{figure}[t]
   \centering
   \includegraphics[scale=0.49]{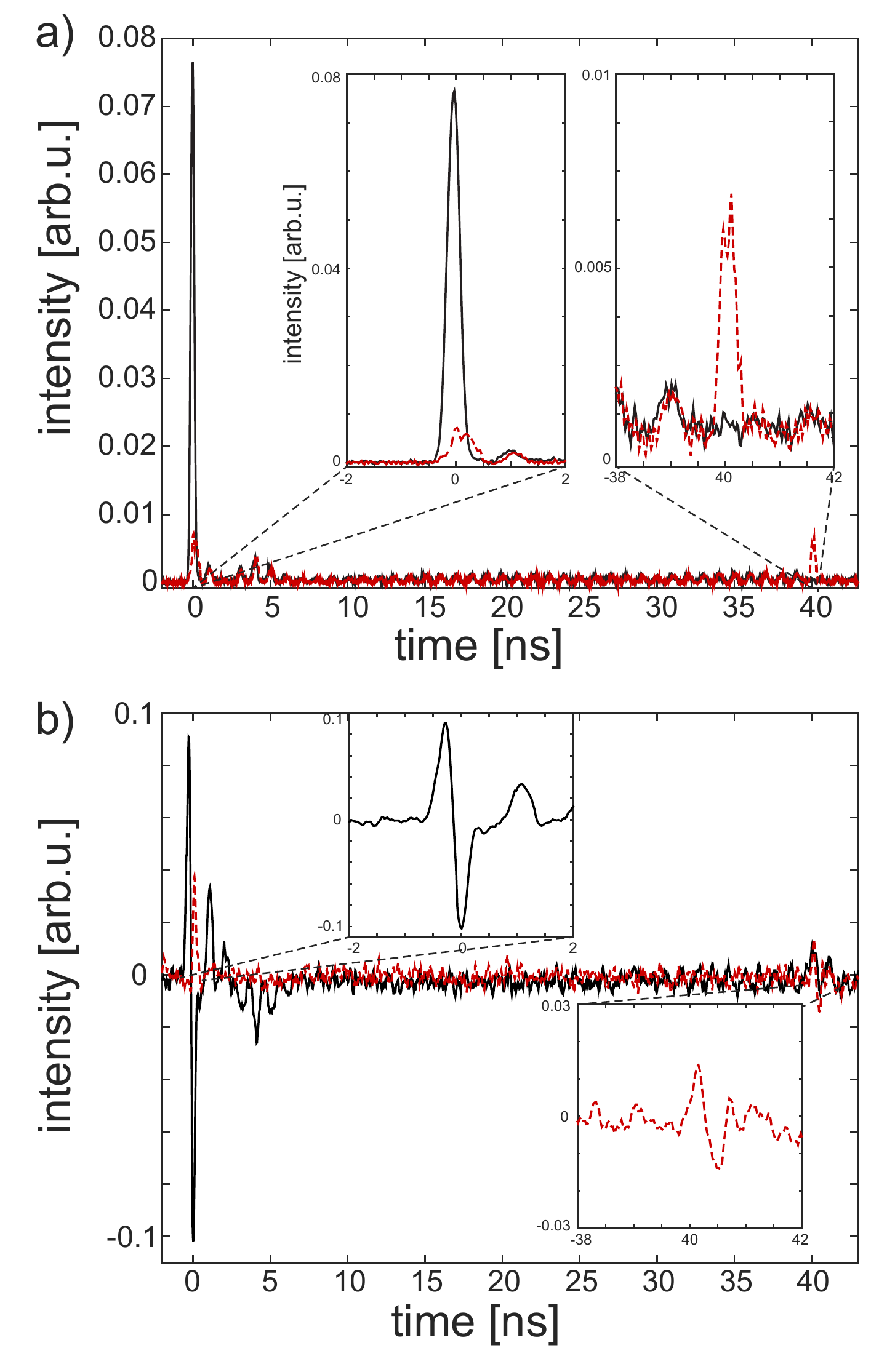}
 \caption{Refreshed memory for read-out after 40\,ns: a) direct detection and b) coherent phase retrieval.}
 \label{fig4}
\end{figure}

The increase of the efficiency allows for a far more important feature which is extending the storage time beyond the intrinsic acoustic lifetime, which so far limited the storage time of the memory to a few nanoseconds. As a proof of principle, we show in Fig. 3 that refreshing the acoustic phonons enables a storage time far beyond the acoustic lifetime of about 10\,ns in As$_2$S$_3$, in this case 40\,ns. In Fig. 3a, an original data pulse (black solid line) is transferred to an acoustic phonon. The latter is refreshed by 39 consecutive refresh pulses which are sub-sequentially filtered out by an optical switch before detection. 40\,ns after the initial data pulse, a read pulse converts the information back to the optical domain and we retrieve our delayed optical pulse (red dashed line). 
In a second measurement we use homodyne detection to show that this process is coherent by storing and retrieving the optical phase. In this experiment, two optical pulses with opposite optical phase ``$0$" and ``$\pi$" are sent into the chip and stored via a counter propagating write pulse. The phase is detected via homodyne detection. After refreshing the memory with 39 pulses, we can detect the phase information after 40\,ns (Fig. 3b). 

In order to illustrate the transfer of energy of the refresh pulses to the acoustic wave, we show the transmitted refresh pulses without the optical switch (Fig. 4). In Fig. 4a, the original data pulse and the refresh pulses are depicted (black solid trace). When switching on the write and read process(red dashed trace), one can see, that the original data is depleted and that the refresh pulses loose energy which is transferred to the coherent acoustic phonons and backscattered pulses at frequency $\omega_{\mathrm{write}}$ = $\omega_{\mathrm{refresh}}$ - $\omega_{\mathrm{acoustic}}$. In Fig. 4b, the refresh pulses have been suppressed by an optical switch. As mentioned, this method can be improved by using the orthogonal polarization, which was not possible in our case, due to high polarization-depended loss of the photonic chips.

\begin{figure}[t]
   \centering
   \includegraphics[scale=0.49]{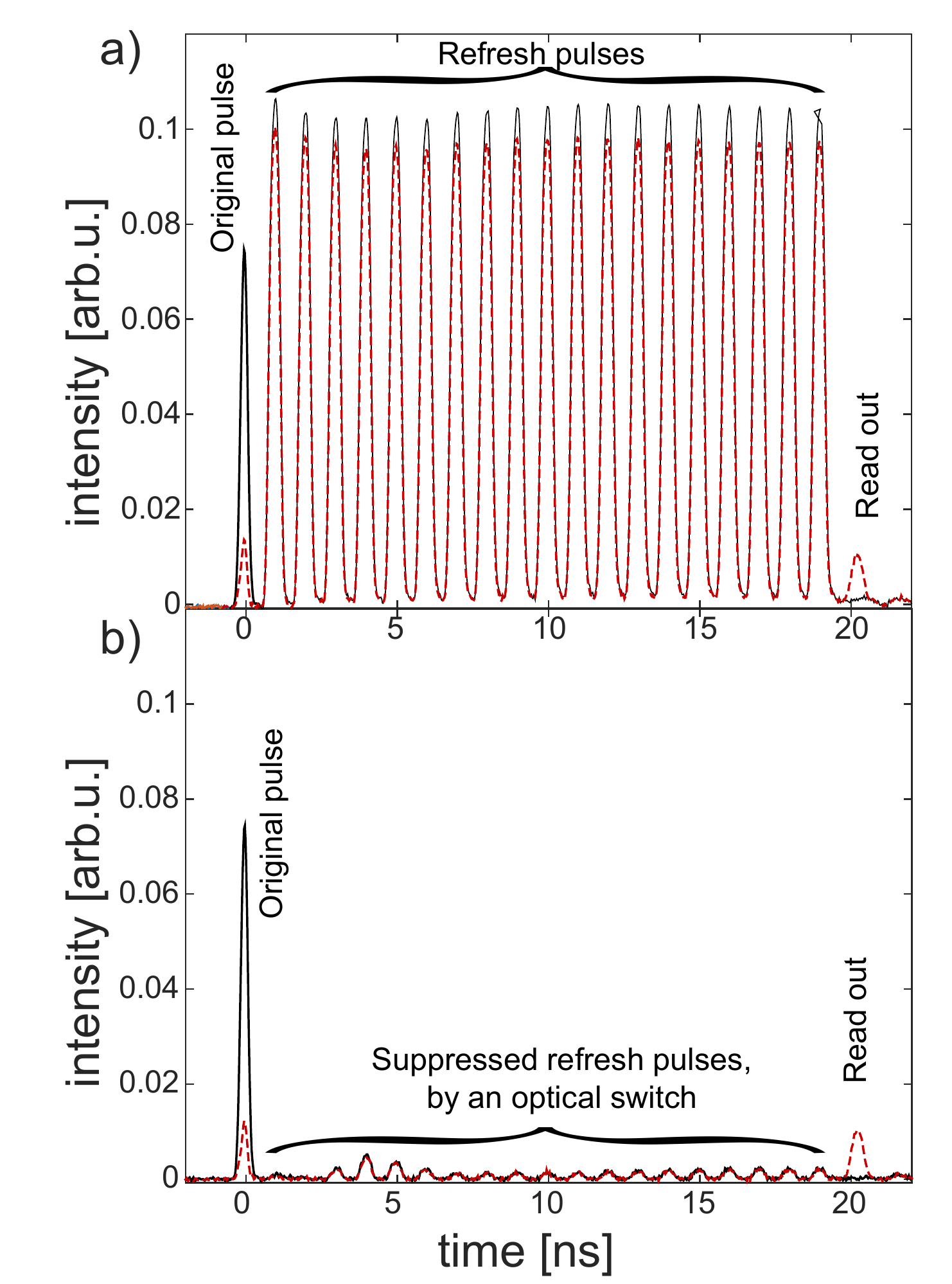}
 \caption{Refreshed memory for storage time of 20\,ns: a) Original data pulse with 19 refresh pulses (black solid line); depleted and retrieved data (red dashed line) with refresh pulses transferring energy to the acoustic phonon; b) Original data (black solid line) and retrieved data (red dashed line) with suppressed refresh pulses by an optical switch.}
 \label{fig3}
\end{figure}

Our experimental setup appeared to be limited by the following pre-dominant factors: first, the extinction ratio of the optical modulators leads to a non-zero background between the optical pulses, which acts as a seed for acoustic phonons that do not hold information. This ultimately limits the detection of the relevant stored information. Second, spontaneous Brillouin scattering can build up, initiated by room temperature phonons and amplified by the refresh pulses. A third limitation is the SNR of the retrieved optical pulse in the photodetection process, limited by the electronic noise of the detector and the oscilloscope. At last, the acoustic dynamic grating broadens with each refresh process due to the convolution with refresh pulse with a finite width, which limits the detectable signal at the photodiode.

While the experimental results demonstrate that the limits of an unrefreshed Brillouin-based memory can be beaten, the question arises how long the storage time can be extended. In other words, how does the SNR evolve over time such that we can still recover the information.
To answer this, we performed a simple analysis of the noise accumulation 
assuming a train of Dirac-shaped refresh pulses (spaced by some time $\tau$) 
chosen such that the acoustic amplitude is kept constant on average.
We decompose the acoustic field into the non-fluctuating excitation (the stored 
pulse including accumulated noise) and a fluctuating field caused by thermal 
excitations.
Both seek to exponentially approach thermal equilibrium with the acoustic
decay constant $\alpha$.
A refresh pulse amplifies both fields, adding a snapshot of the fluctuating
field to the stored pulse.
This effectively ``resets'' the fluctuation field, which is exponentially
repopulated $\sim [1 - \exp(-\alpha t)]$.
As a result, the SNR ratio after $n$ loss-compensating refresh pulses is
\begin{align}
  \text{SNR}_\text{refreshed}(n\tau) \approx 
  \frac{\text{SNR}_\text{initial}}{n [1 - \exp(-\alpha \tau)]}.
\end{align}
This means that the exponential decay $\text{SNR} \simeq \exp(-\alpha t)$ of 
the unrefreshed is transformed into a first order algebraic decay 
$\text{SNR} \simeq t^{-1}$.
This means that refreshing dramatically extends the visibility of stored
information.
For example, doubling the initial SNR doubles the practical read-out time with
refreshing, while it only leads to a constant extension $\ln(2) / \alpha$ 
without refresh.
In reality, the refresh pulses have a finite width and each refresh operation
applies a convolution to the stored data~\cite{Santagiustina2013}.
In our case, this effectively leads to a dispersion-like broadening of the
dynamic grating without imparing the coherence or bandwidth of the signal.
Assuming that this convolution effect can be reverted (e.g. using chirped
pulses \cite{Winful2013a}) and assuming loss-compensating ideal refresh pulses, we 
estimate that it should be possible to maintained data in our system for at 
least $350\,\text{ns}$.
This is based on the apparent SNR of the experiment, which includes 
significant detector noise in addition to the thermal acoustic noise and a 
constant background due to the finite extinction ratio of the modulator.
Therefore, storage times into the microsecond range are within reach.
 

We demonstrated a way to compensate for the intrinsic acoustic decay of a coherent acoustic phonon in a chip-integrated waveguide. This leads to an increase in efficiency of the Brillouin-based memory and importantly allows to overcome the limitation in storage time set by the acoustic lifetime. The acoustic phonons are coherently refreshed allowing the storage and retrieval of the optical phase. This demonstration overcomes the usual constraint of the bandwidth-delay product and paves the way for long phonon-based light storage. Conservative estimation promises storage times into the microsecond regime while conserving the large GHz-bandwidth of the optical pulses. The resulting bandwidth-delay product can therefore exceed the regime of electromagnetically induced transparency (EIT) systems \cite{Longdell2005,Heinze2013} while being fully integrated on a photonic chip. Refreshing the acoustic waves and therefore increasing the efficiency and extending the storage time is relevant for a number of applications such as telecommunication networks, optical interconnects and ultimately may be interesting for quantum communication systems.

\section*{Methods}
\noindent \textbf{Experimental setup.} The laser source, a narrow-linewidth distributed feedback
laser at 1550 nm, is split into the data/refresh and the control (write, read) pulse arm. The data arm is frequency up-shifted by the Brillouin frequency shift Ω via a
single-sideband modulator. The data, refresh and control pulses are imprinted by two intensity modulators
connected to an arbitrary waveform generator. The control pulses are amplified by an
erbium-doped fibre amplifier (EDFA). The amplified write and read pulses pass
through a nonlinear fibre loop which efficiently suppresses any noise or coherent background
present from the laser or amplifier, respectively. It also has the effect that it changes the pulse area to ensure efficient coupling between the data and control pulses. After the loop a second EDFA amplifies the pulses again. Bandpass filters (bandwidth 0.5 nm) in both arms minimise the amplified spontaneous emission from the EDFAs. The refresh pulses are encoded by the same modulator as used for the data pulses. The data/refresh path leads to one side of the photonic chip, the control pulse path to the opposite side. Before detection a tunable narrowband filter ($\approx$ 4\,GHz) is used to stop backreflections from the control pulses. Moreover an optical switch is inserted, to remove the refresh pulses in the detection scheme. The optical switch can be made redundant when using the other polarization for the refresh pulses.\\
\noindent \textbf{Direct detection.} 
For measuring the amplitude, the original and retrieved data are observed by a 12\,GHz photodiode at the oscilloscope.\\
\textbf{Homodyne detection.} 
For measuring the optical phase, the data pulses are interfered with a local oscillator (CW) at the same wavelength. The beat signal is sent to a polarisation beam splitter which sends the paths with different polarization to the two inputs of a balanced photodetector. The polarisation
of the local oscillator and the data pulses are controlled such that the difference signal of both photodiodes of the balanced photodetector is maximised.

\section*{Acknowledgements}
This work was sponsored by the Australian Research Council (ARC) Laureate Fellowship (FL120100029) and the Centre of Excellence program (CUDOS CE110001010). We acknowledge the support of the ANFF ACT. C.W. acknowledges funding from a MULTIPLY fellowship under the Marie Skaodowska-Curie COFUND Action (grant agreement No. 713694).

%
%


\begin{thebibliography}{99}

\bibitem{Safavi-Naeini2011b}
Safavi-Naeini, A.~H. and Painter, O., ``Proposal for an optomechanical traveling wave phonon-photon translator,'' New J. Phys. \textbf{13}, 013017 (2011).

\bibitem{Pant2011}  
Pant, R. et al., ``On-chip stimulated Brillouin scattering,'' Opt. Express \textbf{19}, 8285–8290 (2011).
  
\bibitem{Rakich2013}
Shin,  H., et al., ``Tailorable stimulated Brillouin scattering in nanoscale silicon waveguides,'' Nat. Comm. \textbf{4,} 1944 (2013).
  
\bibitem{Eggleton2013}
Eggleton, B. J., Poulton, C. G. and Pant, R. ``Inducing and harnessing stimulated Brillouin scattering in photonic integrated circuits,'' Adv. Opt. Photonics \textbf{5}, 536–587 (2013).

\bibitem{Beugnot2014}
Beugnot, J.-C., Lebrun, S., Pauliat, G., Maillotte, H., Laude, V. and Sylvestre, T., ``Brillouin light scattering from surface acoustic waves in a subwavelength-diameter optical fibre,'' Nat. Comm. \textbf{5}, 5242 (2014).

\bibitem{VanLaer2015}
 Van Laer, R., Bazin, A., Kuyken, B., Baets, R., Van Thourhout, D., ``Net on-chip Brillouin gain based on suspended silicon nanowires,'' New J. Phys. \textbf{17} 115005 (2015).
  
\bibitem{Balram2015}
Balram, K.~C., Davan{\c{c}}o, M.~I., Song, J.~D., and Srinivasan, K., ``Coherent coupling between radiofrequency, optical and acoustic waves in piezo-optomechanical circuits,'' Nat. Photonics \textbf{10}, 346--352 (2016).
  
\bibitem{Li2015a}
Li, H., Tadesse, S.~A., Liu, Q. and Li, M., ``Nanophotonic cavity optomechanics with propagating acoustic waves at frequencies up to 12 GHz,'' Optica \textbf{2}, 826--831 (2015).

\bibitem{Zhu2007}
Zhu, Z., Gauthier, D. J., and Boyd, R. W., ``Stored light in an optical fiber via stimulated Brillouin scattering,'' Science \textbf{318}, 1748–50 (2007).

\bibitem{Chang2010a}
Chang, D.~E., Safavi-Naeini, A.~H., Hafezi, M. and Painter, O., ``Slowing and stopping light using an optomechanical crystal array,'' New J. Phys. \textbf{13}, 023003 (2011). 

\bibitem{Safavi-Naeini2011}
Safavi-Naeini, A.~H. et~al., ``Electromagnetically induced transparency and slow  light with optomechanics,'' Nature \textbf{472}, 69--73 (2011).
  
\bibitem{Fiore2011}
Fiore, V., Yang, Y., Kuzyk, M.-C., Barbour, R., and Wang, H., ``Storing optical information as a mechanical excitation in a silica optomechanical resonator,'' Phys. Rev. Lett. \textbf{107}, 1–5 (2011).

\bibitem{Jamshidi2012}
Jamshidi, K., Preuler, S., Wiatrek, A., and Schneider, T., ``A review to the all-optical quasi-light storage,'' IEEE Journal on Selected Topics in Quantum Electronics \textbf{18}, 884--890 (2012).

\bibitem{Fiore2013}
Fiore, V., Dong, C., Kuzyk, M.~C. and Wang, H., ``Optomechanical light storage in a silica microresonator,'' Phys. Rev. A \textbf{87}, 1--6 (2013).

\bibitem{Galland2014}
Galland, C., Sangouard, N., Piro, N., Gisin, N., and Kippenberg, T.~J., ``Heralded single-phonon preparation, storage, and readout in cavity optomechanics,'' Phys. Rev. Lett. \textbf{112}, 1--6 (2014).
  
\bibitem{Dong2015}
Dong, C.-H., Shen, Z., Zou, C.-L., Zhang, Y.-L., Fu, W. and Guo, G.-C., ``Brillouin-scattering-induced transparency and non-reciprocal light storage,'' Nature Comm. \textbf{6}, 6193, doi:10.1038/ncomms7193 (2015).

\bibitem{Merklein2017}
Merklein, M., Stiller, B., Vu, K., Madden, S. J. and Eggleton, B. J., ``A chip-integrated coherent photonic-phononic memory,'' Nature Comm. \textbf{8}, 574, doi:10.1038/s41467-017-00717-y (2017).

\bibitem{Merklein2018}
Merklein, M., Stiller, B., and Eggleton, B. J., ``Brillouin based light storage and delay techniques,'' Journal of Optics \textbf{20}, 083003 (2018).

\bibitem{Stiller2019}
Stiller, B., Merklein, M., Poulton, C. G., Vu, K., Ma, P., Madden, S. J., and Eggleton, B. J., ``Crosstalk-free multi-wavelength coherent light storage via Brillouin interaction,'' APL Photonics \textbf{4}, 040802 (2019).

\bibitem{Brillouin1922}
Brillouin, L., ``Diffusion de la lumi\`{e}re par un corps transparent homog\`{e}ne,''
Annals of Physics \textbf{17,} 88--122 (1922).

\bibitem{Bloembergen1965}
Shen, Y.~R., and Bloembergen, N., ``Theory of Stimulated Brillouin and Raman Scattering,''
  Phys. Rev. A \textbf{137,} 1787 (1965).
  
\bibitem{Boyd1990}
Boyd, R., Rzaewski, K., and Narum, P., ``Noise initiation of stimulated Brillouin scattering,'' Phys. Rev. A \textbf{42}, 5514 (1990).

\bibitem{Gaeta1991}
Gaeta, A. L., and Boyd, R. W., ``Stochastic dynamics of stimulated Brillouin scattering in an optical fiber,'' Phys. Rev. A, \textbf{44}, 3205 (1991).

\bibitem{Townes1963}
Garmire, E., Pandarese, F. and Townes, C. H., ``Coherently driven molecular vibrations and light modulation,'' Phys. Rev. Lett. \textbf{11}, 160 (1963).

  
\bibitem{Winful2015}
Dong, M., and Winful, H.~G., ``Area dependence of chirped-pulse stimulated Brillouin scattering: implications for stored light and dynamic gratings,'' Journal of the Optical Society of America B \textbf{32}, 2514 (2015).
  
  \bibitem{Santagiustina2013}
Santagiustina, M., Chin, S., Primerov, N., Ursini, L., and Th{\'{e}}venaz, L., ``All-optical signal processing using dynamic  Brillouin gratings,''
Sci. Rep. \textbf{3}, doi:10.1038/srep01594 (2013).
  
\bibitem{Winful2013a}
Winful, H., ``Chirped Brillouin dynamic gratings for storing and compressing light,'' Opt. Express \textbf{21}, 10039--10047 (2013).
  
\bibitem{Longdell2005}
Longdell, J. J., Fraval, E., Sellars, M. J., and Manson, N. B., ``Stopped Light with Storage Times Greater than One Second Using Electromagnetically Induced Transparency in a Solid,'' Phys. Rev. Lett. \textbf{95}, 063601 (2005).
  
\bibitem{Heinze2013}
Heinze, G., Hubrich, C., and Halfmann,T., ``Stopped Light and Image Storage by Electromagnetically Induced Transparency up to the Regime of One Minute,'' 
Phys. Rev. Lett. \textbf{111}, 033601 (2013).


\end{thebibliography}
\end{document}